\documentclass[review, a4paper, 10pt, oneside, twocolumn, 3p]{elsarticle}

\usepackage{lineno,hyperref}
\usepackage[T1]{fontenc}
\usepackage[utf8]{inputenc}
\usepackage{lmodern}
\usepackage[english]{babel}
\usepackage{graphicx}
\usepackage{float}
\usepackage{url}
\usepackage{multirow}
\usepackage{subfig}
\usepackage{amsmath}
\usepackage{mathtools}
\usepackage{enumitem}
\usepackage{rotating}
\usepackage{resizegather}
\usepackage{textgreek}
\usepackage{gensymb}

\modulolinenumbers[5]

\journal{Journal of Nuclear Instruments and Methods in Physics Research Section A}

\usepackage{siunitx}
\usepackage{xspace}

\newcommand{\NEW}{\mbox{NEXT-White}}

\newcommand{\fig}{figure}

\newcommand{\Fig}{Figure}

\newcommand{\micro}{\ensuremath{\mu}}

\newcommand{\NGA}{\ensuremath{n_\gamma}}

\newcommand{\st}{\ensuremath{S_2}}
\newcommand{\so}{\ensuremath{S_1}}

\newcommand{\NIT}{\ensuremath{N_2}}
\newcommand{\Qbb}{\ensuremath{Q_{\beta\beta}}}

\newcommand{\BI}{\ensuremath{{}^{214}}Bi}

\newcommand{\Kr}[1]{\ensuremath{^{#1}\mathrm{Kr}}\xspace}

\newcommand{\Rb}[1]{\ensuremath{^{#1}\mathrm{Rb}}\xspace}

\DeclareSIUnit\c{\mbox{$c$}}
\DeclareSIUnit\magn{\mbox{$\times$}}
\DeclareSIUnit\min{min}
\DeclareSIUnit\week{week}
\DeclareSIUnit\year{yr}
\DeclareSIUnit\years{years}
\DeclareSIUnit\yr{yr}
\DeclareSIUnit\standard{std}
\DeclareSIUnit\str{sr}
\DeclareSIUnit\ppm{ppm}
\DeclareSIUnit\ppb{ppb}
\DeclareSIUnit\ppt{ppt}
\DeclareSIUnit\pe{PE}
\DeclareSIUnit\spe{SPE}
\DeclareSIUnit\ev{events}
\DeclareSIUnit\ct{counts}
\DeclareSIUnit\neutron{\mbox{$n$}}
\DeclareSIUnit\smp{samples}
\DeclareSIUnit\Sample{S}
\DeclareSIUnit\ch{ch}
\DeclareSIUnit\hit{hit}
\DeclareSIUnit\hits{hits}
\DeclareSIUnit\bin{(\mbox{5-PE}~bin)}
\DeclareSIUnit\sgm{\mbox{$\sigma$}}
\DeclareSIUnit\rms{RMS}
\DeclareSIUnit\keVr{\mbox{keV$_{\rm nr}$}}
\DeclareSIUnit\keVee{\mbox{keV$_{e{\rm e}}$}}
\DeclareSIUnit\ph{photon}
\DeclareSIUnit\pes{pes}
\DeclareSIUnit\el{electrons}
\DeclareSIUnit\pm{PMT}
\DeclareSIUnit\inch{"}
\DeclareSIUnit\bit{bit}
\DeclareSIUnit\sample{samples}
\DeclareSIUnit\barn{barn}
\DeclareSIUnit\bara{bar}
\DeclareSIUnit\barg{barg}
\DeclareSIUnit\mlardepth{\mbox(meter~of~\LAr~depth)}
\DeclareSIUnit\Curie{Ci}
\DeclareSIUnit\psi{psi}
\DeclareSIUnit\parsec{pc}
\DeclareSIUnit\liveday{\mbox{live-days}}
\DeclareSIUnit\days{\mbox{days}}
\DeclareSIUnit\day{\mbox{day}}
\DeclareSIUnit\miles{\mbox{miles}}
\DeclareSIUnit\degreeC{\mbox{$^{\circ}$C}}
\DeclareSIUnit\electron{\mbox{$e^-$}}
\DeclareSIUnit\Euro{\mbox{\euro}}
\DeclareSIUnit\cph{cph}
\DeclareSIUnit\neq{neq}
\DeclareSIUnit\unit{unit}
\DeclareSIUnit\byte{Byte}
\DeclareSIUnit\Bq{\becquerel}

\newcommand{\HPXeEL}{HPXe-EL}

\newcommand{\TPB}{Tetraphenyl Butadiene}
\newcommand{\PDOT}{Poly Ethylenedioxythiophene}
\newcommand{\ITO}{Indium tin oxyde}

\newcommand{\SQRE}{\ensuremath{1/\sqrt{E}}}

\newcommand{\R}{\ensuremath{r}}

\newcommand{\Z}{\ensuremath{z}}

\newcommand{\ResolutionKrFullFourSevenThreeFourWithSystematics}{\SI{4.55 +- 0.01}{\percent}}
\newcommand{\ResolutionKrFullFourSevenThreeFourQbbWithSystematics}{\SI{0.592 +- 0.001}{\percent}}
\newcommand{\ResolutionKrFidFourSevenThreeFourWithSystematics}{\SI{3.88 +- 0.04}{\percent}}
\newcommand{\ResolutionKrFidFourSevenThreeFourQbbWithSystematics}{\SI{0.504 +- 0.005}{\percent}}

\newcommand{\RbLifetime}{\SI{86.2}{\days}}

\newcommand{\KrFidVolumeRRunII}{\SI{150}{\mm}}

\newcommand{\KrFidVolumeZRunII}{\SI{150}{\mm}}

\newcommand{\NewSevenBarPressureRunII}{\SI{7.2}{\bar}}

\newcommand{\NewTpcLength}{\SI{664.5}{\mm}}

\newcommand{\NewTpcDriftLength}{\SI{530.3 +- 2}{\mm}}

\newcommand{\NewNumberOfSiPM}{\num{1792}}

\newcommand{\NewSipmPitch}{\SI{10}{\mm}}

\newcommand{\NewNumberOfPMT}{\num{12}}

\newcommand{\NewCathodeToPMTs}{\SI{130}{\mm}}

\newcommand{\NewPMTSampling}{\SI{25}{\nano\second}}

\newcommand{\NewTpcDiameter}{\SI{454}{\mm}}
\newcommand{\NewFiducialMass}{\SI{5}{\kg}}

\newcommand{\NewPressure}{\SI{15}{\bar}}

\newcommand{\NewPmtEndCapThickness}{\SI{120}{\mm}}

\newcommand{\NewTypePMT}{Hamamatsu R11410-10}
\newcommand{\NewPMTCoverage}{31\%}

\newcommand{\NewBaseEpoxy}{\SI{1.4}{\milli\becquerel\per\kilogram}}
\newcommand{\NewBaseCable}{\SI{46.8}{\milli\becquerel\per\kilogram}}

\newcommand{\NewPMTOperatingVoltage}{\SI{1.23} {\kV}}

\newcommand{\FC}{\ensuremath{\rm f_{cutoff}}}

\begin{document}

\setlength{\parskip}{4mm}
\sloppy \lefthyphenmin=1000

\begin{frontmatter}

\title{The electronics of the energy plane of the NEXT-White detector}

\author[a]{V. \'Alvarez\corref{mycorrespondingauthorprime}}
 \ead{vicente.alvarez@ific.uv.es} 
 \cortext[mycorrespondingauthorprime]{Corresponding author.}
\author[b]{V. Herrero-Bosch}
\author[b]{R. Esteve}
\author[a]{A. Laing}
\author[a]{J. Rodr\'iguez}
\author[a]{M. Querol}
\author[c]{F. Monrabal}
\author[b]{J. F. Toledo}
\author[e,d,a]{J.J. G\'omez-Cadenas}

\address[a]{Instituto de F\'isica Corpuscular (IFIC), CSIC \& Universitat de Val\`encia\\ 
Calle Catedr\'atico Jos\'e Beltr\'an, 2, 46980 Paterna, Valencia, Spain}
\address[b]{Instituto de Instrumentaci\'on para Imagen Molecular (I3M), Centro Mixto CSIC - Universitat Polit\`ecnica de Val\`encia\\ 
Camino de Vera s/n, 46022 Valencia, Spain}
\address[c]{
 Department of Physics, University of Texas at Arlington \\ 
Arlington, Texas 76019, USA}
\address[d]{Donostia International Physics Center (DIPC)\\ 
Paseo Manuel Lardizabal 4, 20018 Donostia-San Sebastian.}
\address[e]{IKERBASQUE, Basque Foundation for Science, 48013 Bilbao, Spain.}

\begin{abstract}

This paper describes the electronics of \NEW\ (NEW) detector PMT plane, a high pressure xenon TPC with electroluminescent amplification (\HPXeEL) currently operating at the Laboratorio Subterr\'aneo de Canfranc (LSC) in Huesca, Spain. 
In \NEW\ the energy of the event is measured by a plane of photomultipliers (PMTs) located behind a transparent cathode. 
The PMTs are \NewTypePMT\ chosen due to their low radioactivity. 
The electronics have been designed and implemented to fulfill strict requirements: an overall energy resolution below 1\% and a radiopurity budget of $\SI{20}{\milli\becquerel\per\unit}$ in the chain of \BI\ . All the components and materials have been carefully screened to assure a low radioactivity level and at the same time meet the required front-end electronics specifications. 
In order to reduce low frequency noise effects and enhance detector safety a grounded cathode connection has been used for the PMTs. This implies an AC-coupled readout and baseline variations in the PMT signals. A detailed description of the electronics and a novel approach based on a digital baseline restoration to obtain a linear response and handle AC coupling effects is presented. The final PMT channel design has been characterized with linearity better than $0.4\%$ and noise below $0.4mV$.

\end{abstract}

\begin{keyword}

\texttt{Calorimetry \sep Front-end electronics \sep  Digital Baseline Restoration}

\end{keyword}
\end{frontmatter}

%\linenumbers

\section{\label{sec:introduction}Introduction}

The NEXT program is developing the technology of high-pressure xenon gas Time Projection Chambers (TPCs) with electroluminescent amplification (\HPXeEL) for neutrinoless double beta decay searches (\Qbb) \cite{Nygren:2009zz, Alvarez:2011my, Alvarez:2012haa, Gomez-Cadenas:2013lta, Martin-Albo:2015rhw}. 
% The \NEW\footnote{Named after Prof.~James White, our late mentor and friend.} detector implements the second phase of the program. \NEW\ is a $\sim$ 1:2 scale detector of NEXT-100. The TPC has a length of \NewTpcLength\ and a diameter of \NewAnodePlateDiameter\, a factor of two the dimensions of the NEXT-100 TPC, 100 kg \HPXeEL\ detector, which constitutes the third phase of the program and is foreseen to start operations in 2019. 
% while the NEXT-100 TPC has a length of \NextTpcLength\ and a diameter of  \NextTpcDiameter), a 100 kg \HPXeEL\ detector, which constitutes the third phase of the program and is foreseen to start operations in 2019. 
\NEW\ has been running successfully since October 2016 at Laboratorio Subterr\'aneo de Canfranc (LSC). Its purpose is to validate the \HPXeEL\ technology in a large-scale radiopure detector.
 
 This paper describes the front-end electronics of the plane of photomultipliers (PMTs) used to measure the energy of the events in \NEW, the so-called, {\em energy plane}. It shows a complete system starting at a PMT with a custom designed polarized base whose output signal is conditioned by a custom made front-end board (FEE) and then digitized by a sampling ADC with an on-board signal processing. The DAQ uses a 12-bit ADC with an input range of 2V (LSB = 0.49 mV) sampling at 40 MHz. This acquisition system is shared with other subsystems such as the Tracking Plane and therefore must meet the requirements of all them. The acquisition system is explained in more detail in the section \ref{sec.atca}. Once digitized and framed, this produces approximately 10 MByte/s for the 12 PMTs at a 10 Hz rate. On line digital signal processing required to properly recover PMT signals can also be carried out at the acquisition system level. The main reference specification of EP will be its energy resolution which must be close to the intrinsic energy resolution of the detector itself in order not to excessively degrade its performance. As described in \cite{Martinez-Lema:2018ibw} the Fano factor based intrinsic resolution is 0.3\%. 
 \par The organization is as follows. Section \ref{sec.new} summarizes the main features  of the detector. Section \ref{sec:FEE} presents the front-end electronics (FEE) and Section \ref{Digi_base_res} introduces a novel digital baseline restoration algorithm. Conclusions are stated in section \ref{sec:conclu}.

\section{The \NEW\ detector}
\label{sec.new}

\begin{figure*}[!htb]
  \begin{center}
   	 \includegraphics[width=1\textwidth]{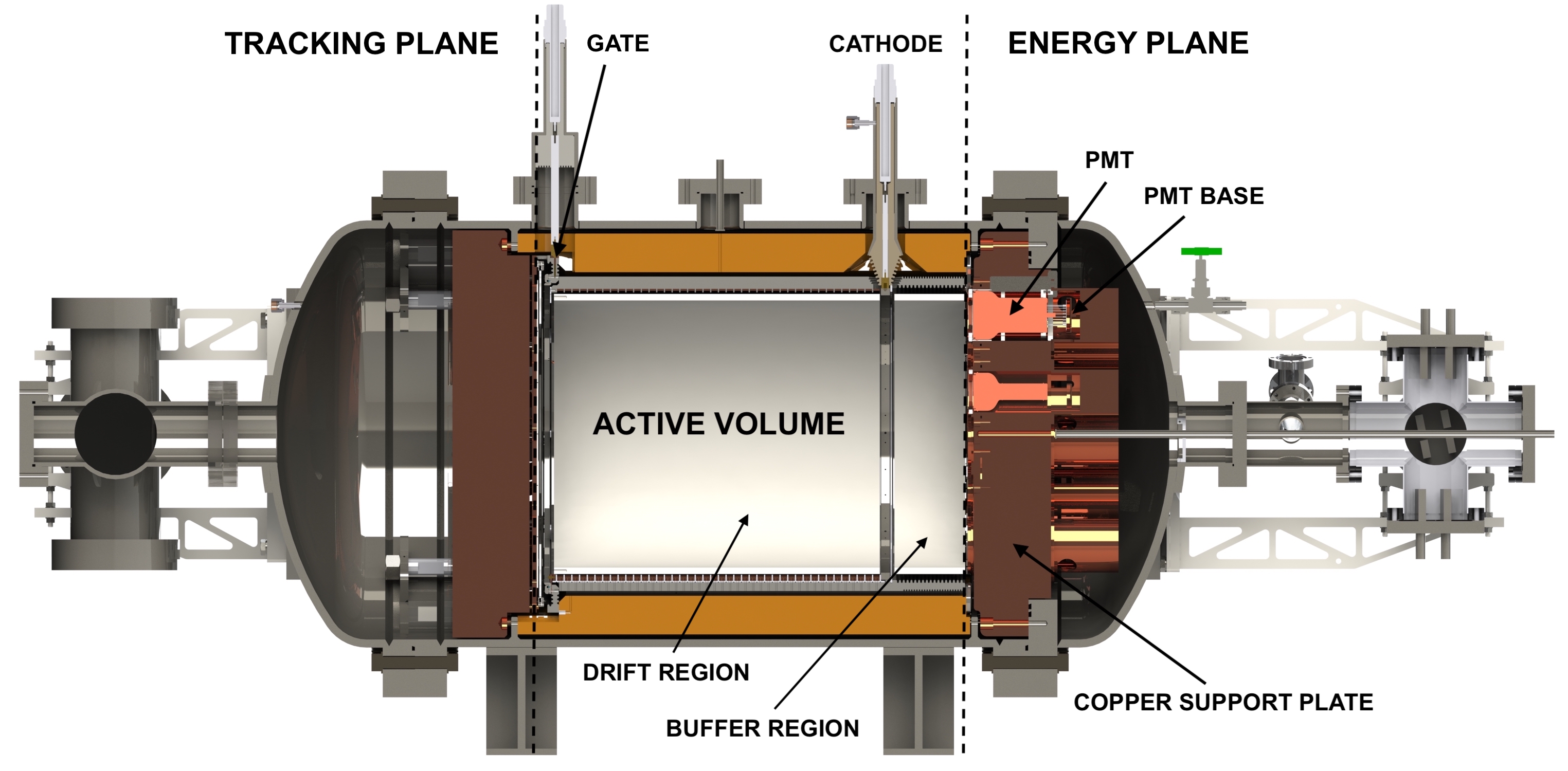}    
   	 \caption{The \NEW\ detector}
   	 \label{fig:NEW}
  \end{center}
\end{figure*}

\Fig\ \ref{fig:NEW} shows a drawing of the \NEW\ detector~\cite{Monrabal:2018xlr}. The main components are:

{\em A Time Projection Chamber (TPC)}, that defines the detector fiducial volume. The TPC has a total length of  \NewTpcLength, a drift length of \NewTpcDriftLength\ and  \NewTpcDiameter\ diameter. It contains \NewFiducialMass\ of xenon mass in the active volume at \NewPressure. Structurally the TPC consists of:  a field cage; two transparent grids, one at the cathode and one at the end of the drift region (called the gate); a quartz plate coated with a thin layer of \ITO\ (ITO), a conductor material used to properly define the anode voltage on the surface of the plate, and a thin layer of \TPB\ (TPB), commonly used in noble gas detectors to shift VUV light to the visible spectrum \cite{Gehman:2011xm}; and two high voltage feedthroughs (HVFT), which are used to set the voltages at the cathode and the gate. Functionally, the TPC includes the drift region which defines the fiducial mass of the detector, the buffer region, needed to smoothly degrade the high voltage at the cathode to ground and the electroluminescent (EL) region, where each drift electron is amplified. Each electron entering the EL region is accelerated by the higher electric field. The accelerated electrons collide with the gas atoms exciting them and producing scintillation light emitted in VUV in the case of xenon. The number of photons emitted, the optical gain \NGA, depends on the electric field in the EL region, the gas pressure and the width of the region and, in NEW, is typically set to 1000 photons per electron. %\NextOpticalGain.
%emits a number \NGA\ of VUV photons as it crosses the EL region; \NGA\ also called {\em optical gain}, depends on the EL electric field and the pressure and is typically set to \NextOpticalGain.

{\em A tracking plane} is located behind the anode and equipped with \NewNumberOfSiPM\ SiPMs SensL series-C SiPMs distributed at a pitch of \NewSipmPitch. The spaces between SiPMs are covered by a thin (2 mm) teflon layer that increases the total anode reflectivity and enhances light collection by the PMTs. 

\begin{figure}[H]
	\centering
	\includegraphics[angle=0, width=0.45\textwidth]{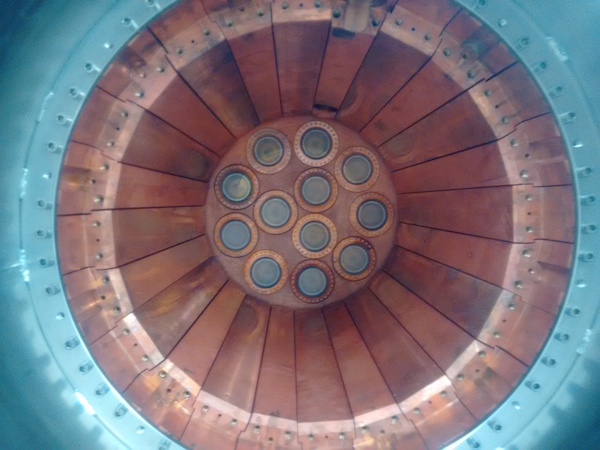}
	\caption{The energy plane viewed from the anode, showing the sapphire windows coated with TPB.}
	\label{EP_plane}
	\vfill
	\includegraphics[angle=0, width=0.45\textwidth]{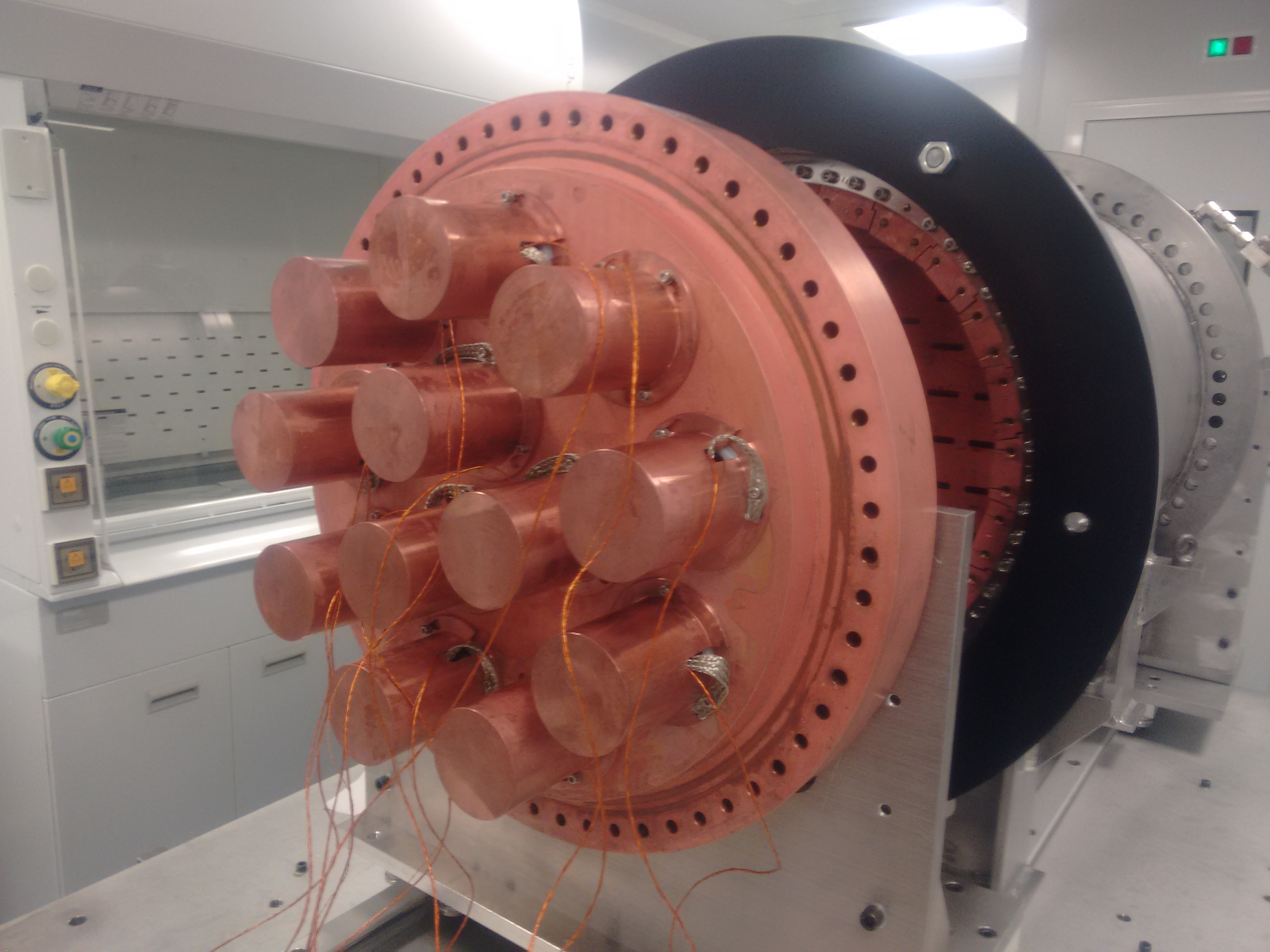}
	\caption{The copper hats that protect the PMTs and shield from external radiation.}
	\label{fig.EP}
\end{figure}

{\em An energy plane} (\fig\ \ref{EP_plane} and \fig\ \ref{fig.EP}) equipped with \NewNumberOfPMT\ \NewTypePMT\  PMTs is located \NewCathodeToPMTs\ behind the cathode and covers \NewPMTCoverage\ of its surface area. This level of coverage was chosen as a compromise between the need to collect as much light as possible for physics measurements and the need to minimize the number of sensors to reduce cost, technical complexity and radioactivity.  The selected model  is a \SI{3}{\inch} PMT specially developed for \mbox{low-background} operation. The operational gain is $2.5 \times 10^6$, it is equipped with a synthetic fused silica window, 12 amplifying stages and a bialkali photocatode with a quantum efficiency of 26\% at 175 nm. The PMTs receive high voltage and have their signal extracted via kapton twisted cables connected to a feedthrough in the torispherical head of the pressure vessel. Distribution of signal and supply at each individual PMT is done by means of a Kapton circuit board (PMT base) which is covered with a copper cap and filled with radiopure epoxy. %(\fig\ \ref{fig:7C} and \fig\ \ref{fig:7C_layout}).
 PMT bases are connected to the copper support plate described below to allow generated heat to be dissipated under vacuum conditions. The thermal connection has been done via copper braid that connects the copper cap in the base with the copper hat(\fig\ \ref{fig.EP}).

Since the \NewTypePMT\ can not operate at high pressure, they are separated from the active volume by a \NewPmtEndCapThickness\ copper plate. The hermetically sealed PMT region is then filled with one atmosphere of \NIT. The PMTs are coupled to the xenon gas volume through \NewNumberOfPMT\ holes, machined in the copper plate and closed with sapphire windows. These windows are coated with a resistive (and very transparent) compound \PDOT\ (PDOT) \cite{sigma:pedot} in order to define a ground while at the same time avoiding sharp electric field components near the PMT windows and a thin layer of TPB (\fig\ \ref{EP_plane}).

\section{Front-end electronics of the \NEW\ energy plane}
\label{sec:FEE} 

One of the most distinctive features of an \HPXeEL\ is its
excellent energy resolution, with an intrinsic limit (given by the Fano factor) of about 0.3\% FWHM at the xenon endpoint \Qbb\ (2.458 MeV). The NEXT prototypes \cite{Alvarez:2012hh} as well as the initial results of \NEW\ \cite{Martinez-Lema:2018ibw} have also demonstrated an energy resolution for point-like particles (Krypton X-rays) which extrapolates to 0.5\% at \Qbb\ (assuming naive scaling $\sigma_{\Qbb} \sim \sigma_{Kr}/\sqrt{E}$) and to better than $\sim 0.7$ \% FWHM for extended tracks. 

The electronics for the \NEW\ energy plane was complete custom developed by us and has been designed to preserve the intrinsic resolution characterizing an \HPXeEL, as well as to minimize the radioactive budget. For the latter purpose, all the components and materials have been carefully screened. A grounded cathode has been chosen in order to simplify structural design removing the need of a housing isolation. Also a remarkable reduction of low frequency noise effects has been achieved by using this connection strategy.

\begin{figure*}[h]
\centering
  \includegraphics[width=\textwidth]{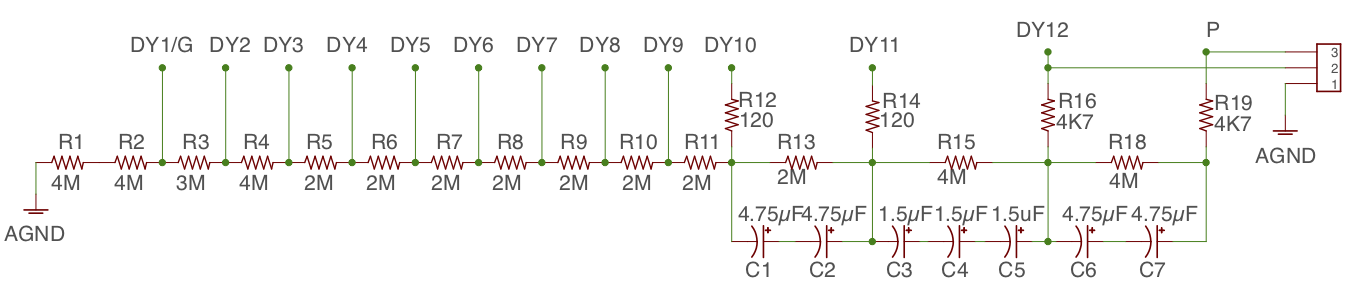} \\
\caption{\NEW\ PMT base circuit.}
\label{fig:7C}
\end{figure*}

\vspace{2mm}

The design has been optimized to preserve linearity over the whole dynamic range from 0.1 photons/ns to 1 photons/ns for light pulses longer than \SI{150}{\micro\second} . This is a crucial feature, in particular to measure the energy of high-energy electrons which result in long tracks in the chamber (and therefore in signals as long as \SI{150}{\micro\second}). To cope with those long signals, the base circuit must provide enough charge for the PMT with negligible change in dynode-to-dynode voltage for a maximum average current of 0.1 mA \cite{HamamatsuPMTs} . Changes in these voltages introduce a time varying gain resulting in a nonlinear distortion mechanism that has a strong effect on PMT linearity.

\subsection{PMT base circuit}

\begin{figure}[H]
	\centering
	\includegraphics[width=0.4\textwidth]{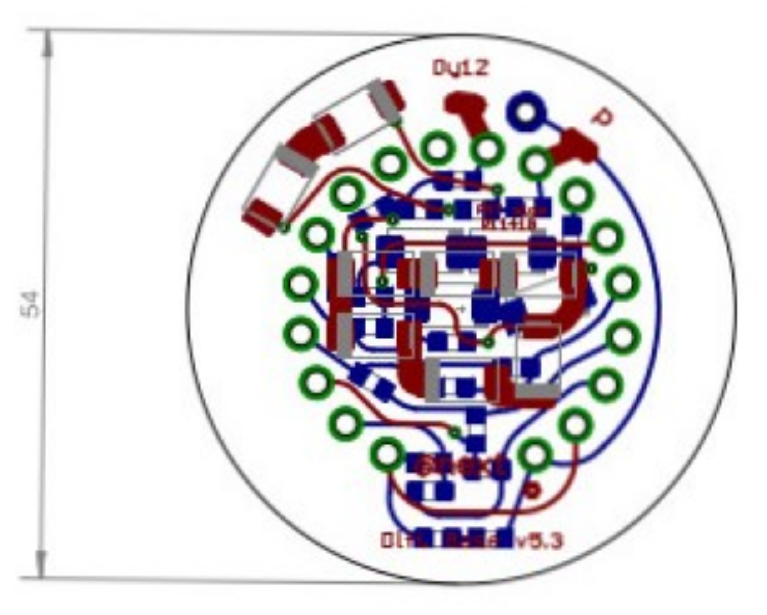}
	\caption{Base layout.} 
	\label{fig:7C_layout}
\end{figure}

\begin{figure}[H]
	\centering
	\includegraphics[width=0.4\textwidth]{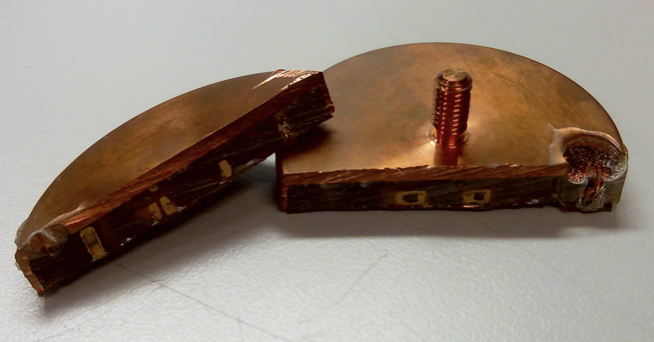}
	\caption{Copper cap covering the PMT base.} 
	\label{fig:7C_base}
	\vfill
	\includegraphics[width=0.5\textwidth]{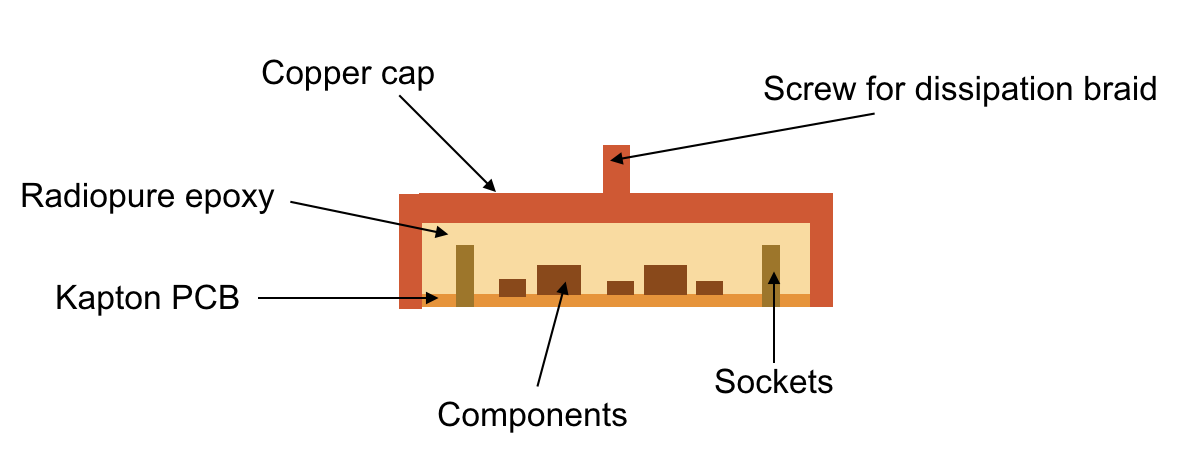}
	\caption{Scheme of the PMT base.}
	\label{fig:7C_base_sch}
\end{figure}

The PMT base is a passive circuit involving 19 resistors of different electrical resistance, 7 tantalum capacitors (3 having a capacitance of \SI{1.5}{\micro\farad} and 4 with  \SI{4.7}{\micro\farad}), 18 pin receptacles soldered on a kapton circuit board (\fig\ \ref{fig:7C} and \fig\ \ref{fig:7C_layout}) covered with epoxy \cite{Araldite2011} to avoid dielectric breakdown in moderate vacuum or in a atmosphere of \NIT, and to improve thermal contact with the copper cap (\fig\ \ref{fig:7C_base} and \fig\ \ref{fig:7C_base_sch}). Critical parameters for the PMT base circuit are the operating voltage (\NewPMTOperatingVoltage) and the corresponding PMT gain ($2.5 \times 10^6$).

\par The voltage applied to the PMT has been established through a trade-off between single photon detection capability and manufacturer recommended high voltage range for best linearity. In order to achieve enough SNR, FEE noise must be reduced which limits FEE maximum gain thus requiring a higher PMT gain. A test bench for linearity measurements has been custom developed to optimize PMT base design and test energy plane detectors under \NEW\ energy plane conditions. Component nominal values and base topology will be explained along the next section.

\subsection {Linearity test bench, component sizing and measurements}

\par
\begin{figure}[H]
	\centering
	\includegraphics[width=0.5\textwidth]{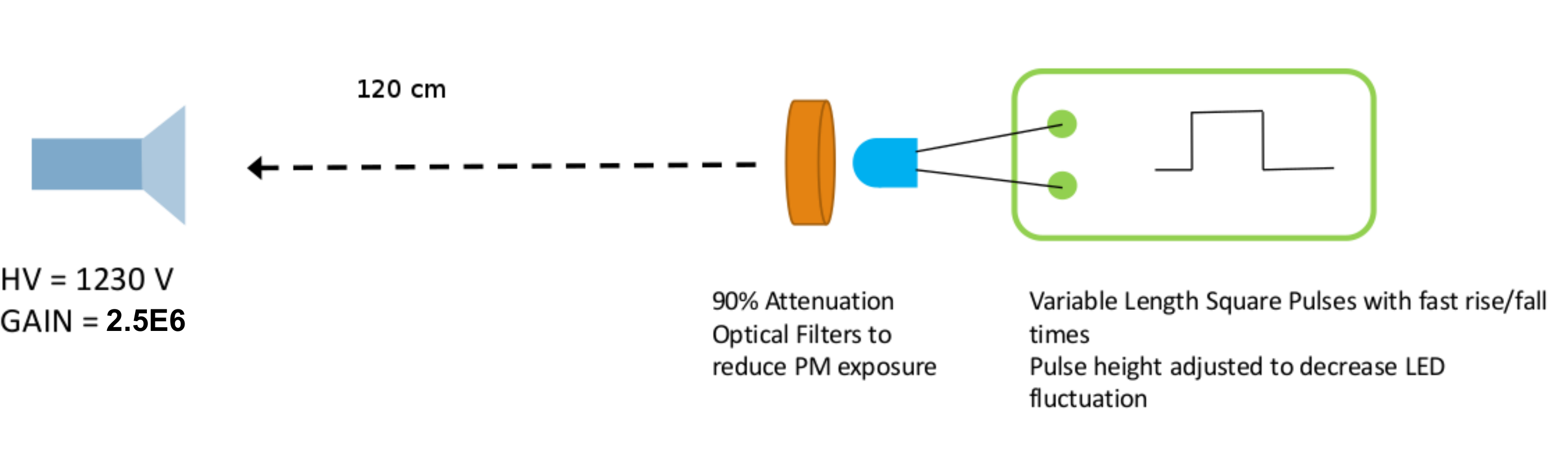}
	\caption{LED based PM+FEE test bench for linearity measurements}
	\label{fig:LED_testbench}
\end{figure}

 The test bench will include not only PMT (with its base) but also the associated FEE due to the special characteristics of the signals captured and the required linearity (\fig\ \ref{fig:LED_testbench}).  The dynamic range of each energy plane channel must cover from a single photon up $100 \times 10^3$ photons. Also the signal length variability is a challenging specification since the PMT must withstand signals from a few microseconds up to 150 $m$s.

%\begin{figure}[!H]
%	\begin{center}
%		\includegraphics[width=0.5\textwidth]{pool/imgEP/LED_testbench.pdf}
%		\caption{LED based PM+FEE test bench for linearity measurements}
%		\label{fig:LED_testbench}
%	\end{center}
%\end{figure}

\par The linearity test bench makes use of a signal generator which introduces a voltage square pulse with very short rise and fall edges into a 435 nm  LED. The number of photons generated is proportional to the length of the pulse. In order to avoid LED switching effects the minimum pulse length is 30 \micro s (the switching time of the LED is in the order of 10 ns).

\par The proposed test bench is a complete system with the base and a grounded anode connected R11410-10 PMT. In order to reduce the effect of the LED fluctuations every measurement has been averaged using 200 samples. Also a set of 90\% attenuation optical filters was used directly coupled to the LED.

\par
\begin{figure*}[!htbp]
  \begin{center}
    \includegraphics[width=1.1\textwidth, angle=0]{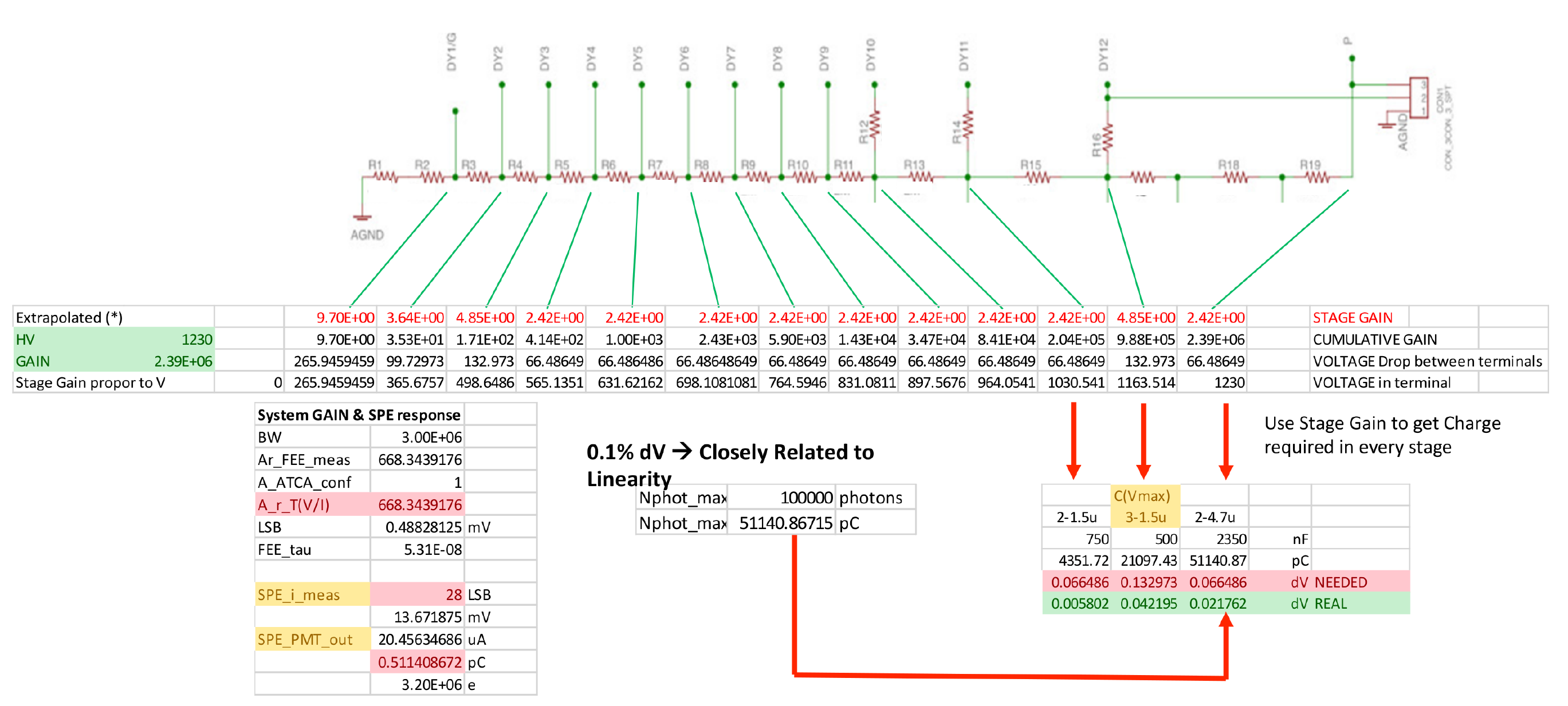}
    \caption{Base Circuit. Capacitor sizing procedure}
    \label{fig:Base_Sizing}
  \end{center}
\end{figure*}

 The capacitors of the design have been computed based on the algorithm shown in \fig\ \ref{fig:Base_Sizing}. The main goal is that the maximum signal in terms of photons introduces a voltage drop below the 0.1\% of the voltage between dynodes. Since each stage increases the gain by a given factor (based on the geometry and characteristics of the PMT) the corner cases for this voltage drop will be located at the last stages where the amount of charge to be delivered by the PMT is higher. Therefore starting from the last PMT stage, a certain number of stages will be decoupled with capacitors.

\par In order to meet the radiopurity limits \cite{Martin-Albo:2015rhw} a first proposal based on a minimum 2 stage coverage (5 capacitors to withstand the high voltage between terminals) has been tested. Although linearity graphs show a quite linear response there are some pulse lengths that fail in the fit. 
 
The output signal doesn't show a typical overload effect which would appear due to a usual fall in the gain value for high number of photons. This behavior is related to a distortion mechanism that appears due to a charge redistribution in the base capacitors. When a capacitor gets depleted, it drains charge from the previous one. When the charge drained by the last capacitor is too high, voltage drop in the previous stage affects the overall gain. The length of the signal pulse able to generate this distortion depends on the time constants of each stage and may change with voltage distribution among stages.

\par
\begin{figure}
	\includegraphics[width=0.4\textwidth]{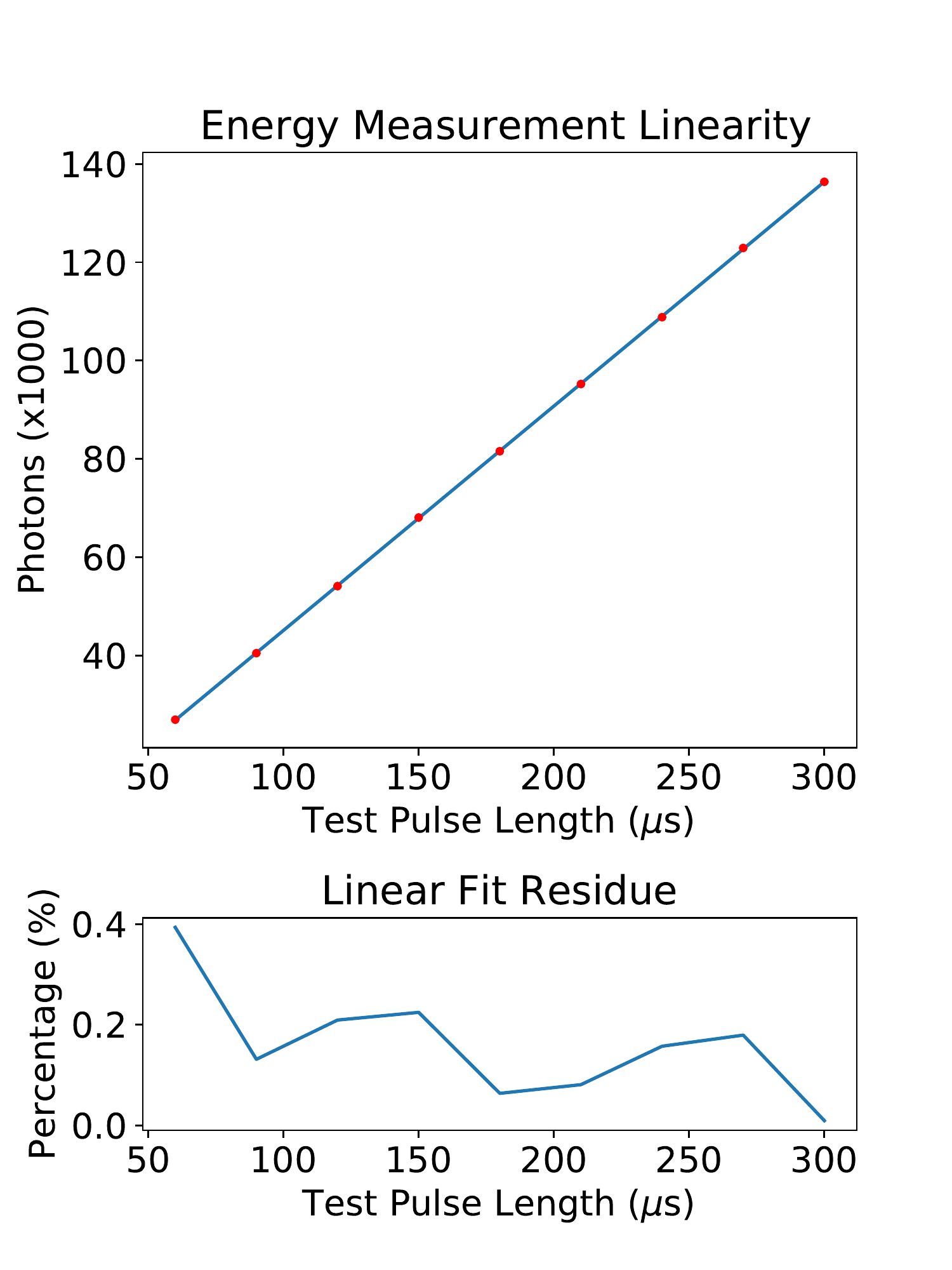}
	\caption{Fit to the response of the PMT plus base (0.5 photons/ns). }
	\label{fig:7C_lin}
\end{figure}

 The measured non-linearity \footnote{Measured as the worst case deviation from the linear fit.} is 1.95\% for 100k pe. Finally this means that a better capacitor coverage is required so a 3 stage coverage was chosen (\fig\ \ref{fig:7C} and \fig\ \ref{fig:Base_Sizing}). This circuit shows virtually no time distortion. It has been designed to optimize the trade off  between the conflicting targets of preserving the signal linearity (\fig\ \ref{fig:7C_lin}) and keeping the radioactive budget to a minimum. The difficulty to achieve both goals simultaneously resides in the fact that the stringent requirements to keep a linear response requires the use of capacitors to hold the charge in the latest amplification stages, where the gain is very large. The linearity fit gives a $0.38\%$ maximum error of linearity for an input of \num{140E+3} photons which exceeds the initial requirements for the maximum charge expected per time bin in the PMTs.

\par The resistor chain value ratios have been recommended by the PMT manufacturer and the final value of the resistors have been adjusted to reduce power dissipation. A power of \SI{40}{\milli\watt} results in a stable temperature of \SI{30}{\degreeC} for a \SI{21}{\degreeC} ambient temperature. This introduces no disturbances in the rest of the detector. The number of capacitors ---which are the most radioactive elements in the base-- have been kept to a bare minimum. They are only introduced in the last three amplification stages where the amount of charge to be delivered by the PMT is higher and  the dynodes voltage needs to be held with the help of capacitors. Tantalum polarized capacitors have been chosen due to their high nominal capacitance and they are the most radiopure ones we have found. Groups of serial connected capacitors were used in order to withstand high voltage between dynodes since they have a maximum voltage lower that the value that we need. 

\par Finally, PMT base circuit radioactivity has been measured as four times larger than that of the PMT itself, as shown, for \BI\ activity in table \ref{tab.PMTBudget} (see ~\cite{Cebrian:2017jzb} for a thorough discussion). Notice, however, that the radioactivity injected in \NEW\ by the base is partially shielded by the copper shield, and contributes roughly the same to the final background count as the PMTs. 

\begin{table}
\caption{Radioactive \BI\ budget of \NEW\ PMTs and base circuits.}
\begin{center}
\begin{tabular}{|c|c|}
\hline
Component & \BI\ activity per unit\\
\hline\hline
PMT &  \SI{0.35}{\milli\becquerel}\\
\hline
Base & \\ 
Capacitors 1.5 uF & \SI{72}{\micro\becquerel} \\
Capacitors 4.7 uF & \SI{123}{\micro\becquerel}\\
Finechem resistors & \SI{4.1}{\micro\becquerel}\\
KOA RS resistors & \SI{7.7}{\micro\becquerel}\\
Pin receptacles & \SI{1.1}{\micro\becquerel} \\
Araldite epoxy & \NewBaseEpoxy\\
Kapton-Cu cable & \NewBaseCable\\
Kapton substrate & \SI{23}{\micro\becquerel}\\
Copper cap & \SI{12}{\micro\becquerel}\\
\hline
Total base  & \SI{1.2}{\milli\becquerel} \\ 
\hline
\end{tabular}
\end{center}
\label{tab.PMTBudget}
\end{table}%

\subsection{Grounded Cathode PMT connection and its consequences}

\begin{figure}[!t]
\centering
\includegraphics[width=0.5\textwidth]{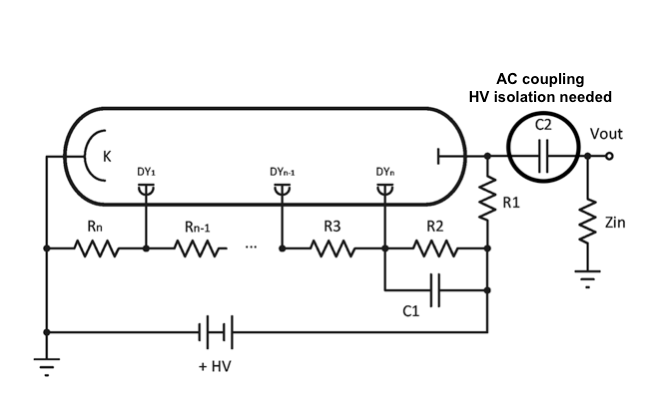}
\caption{Grounded Cathode PMT connection scheme.}
\label{fig:grounded_cathode}
\end{figure}

In \NEW\ the PMTs photocathodes (and thus their bodies) are connected to ground, while  the anode is set at the operating voltage of the PMT (\NewPMTOperatingVoltage).
This solution simplifies the detector mechanics and enhances safety (the alternative, with the cathode and the PMT body at high voltage would have required the insulation of each PMT from ground). In exchange, the anode output needs to be AC coupled through an isolation capacitor, as shown in  \fig\ \ref{fig:grounded_cathode}, since the anode DC voltage equals the high voltage being applied to the PMT. Although a differential transmission line (TX line) will be used in the final implementation, single ended equivalents will be used in the following explanations for the sake of simplicity. As a consequence $Z_{in}$ is related to single ended input equivalent impedance of the amplifier which will be translated into a differential one in the final design.

\begin{figure}[!t]
\centering
\includegraphics[width=0.35\textwidth]{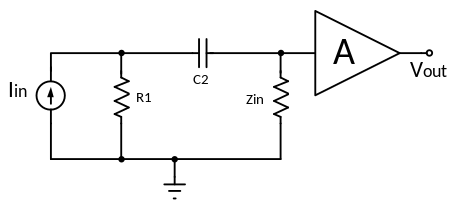} 
\caption{Grounded Cathode Equivalent Circuit.}
\label{fig:GC_equivalent_circuit}
\end{figure}

An AC coupling scheme creates, to first order, a high pass filter (HPF).  \Fig\ \ref{fig:GC_equivalent_circuit} shows the electrical equivalent circuit of the connection scheme, modeling the PMT as a simple current source. The Laplace transfer function of the filter (\fig\ \ref{fig:GC_equivalent_circuit}) is given by 
equation \ref{eq.hpf}:

\begin{equation}
\frac{v_O}{i_I}=A\frac{Z_{in}R_1}{Z_{in}+R_1}\frac{(R_1+Z_{in})C_2s}{1+(R_1+Z_{in})C_2s}
\label{eq.hpf}
\end{equation}
where R$_1$~ is the high value resistor which mainly defines the filter along with the decoupling capacitor C$_2s$. 

The differential line termination (120$\Omega$) will be implemented as twice that of an equivalent single-ended line with a $Z_{in}$ of 60$\Omega$. This option also allows to terminate the common mode signal that travels along the transmission line and is expected to be quite high since a fully differential transmission is not being used.  

The filter blocks the DC component and attenuates frequency components below the cutoff frequency \FC\ , defined as:

\begin{equation}
\FC = \frac{1}{(R_1+Z_{in})C_2 \cdot 2\pi}
\label{eq.hpf2}
\end{equation}

\begin{figure}[!t]
	\centering
	\includegraphics[width=0.5\textwidth]{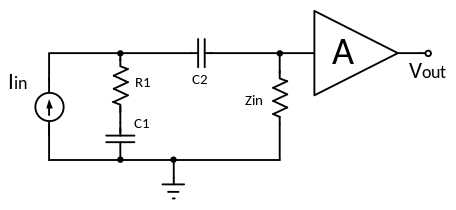}
	\caption{\small A more sophisticated model of the FEE + PMT base.}
	\label{fig:FEE_PMT}
\end{figure}

The output pulse after the filter is the derivative of the input pulse, and is characterized by a null total area. Since the energy is proportional to the area of the input pulse, it follows that a deconvolution (or baseline restoration) algorithm must be applied to the output pulse in order to recover the input pulse area and thus measure the energy. Also, since the intrinsic energy resolution in \NEW\ is very good the error introduced by the deconvolution algorithm must be, at most, a fraction per mil.

Indeed, the simple model of the PMTs as an ideal current source is insufficient given the accuracy required and must be refined in a number of ways. 
For example, the effect of the charge exchange between the capacitors of the PMT base and the coupling capacitor, must be included in the model (\fig\  \ref{fig:FEE_PMT}). The resulting filter is then more complicated than a simple HPF, but can still be accurately described as a pole/zero combination obtained from the equivalent circuit shown in \fig\ \ref{fig:FEE_PMT}. The filter follows the equation: 

\begin{equation}
\frac{v_O}{i_I} =A\frac{Z_{in}}{(1+\frac{C_1}{C_2})}\frac{1+R_1C_1s}{1+\frac{(R_1+Z_{in})C_1}{(1+\frac{C1}{C2})}s} \label{eq:full_fee}
\end{equation}

%\begin{figure}[!htbp]
%	\centering
%	\subfloat[Effect in a square pulse.]{\includegraphics[width=0.45\textwidth]{pool/imgEP/pulse2.pdf}}
%	\vfill
%	\subfloat[Effect in a train of square pulses.]{\includegraphics[width=0.45\textwidth]{pool/imgEP/pulses2.pdf}}
%	\caption{The effect of a HPF.}
%	\label{fig:pulses}
%\end{figure}

\

\begin{figure*}[!htbp]
	\centering
		\includegraphics[width=\textwidth]{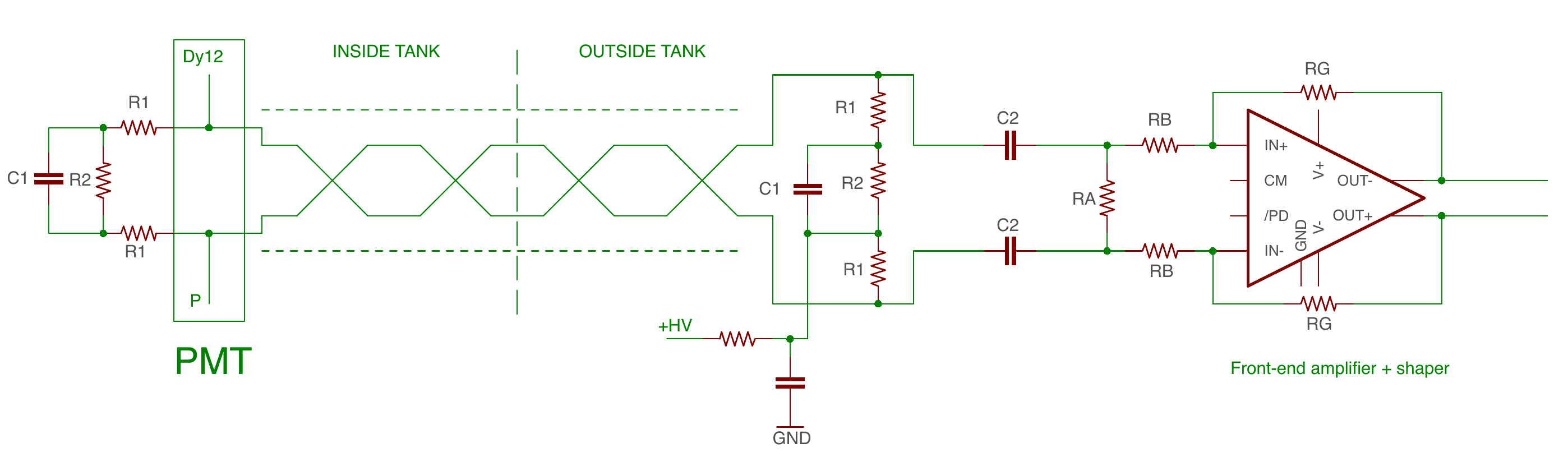}
		\caption {FEE Scheme.} 
		\label{fig:FEE_scheme}
\end{figure*}

Final custom made design is shown in \fig\ \ref{fig:FEE_scheme}. A pseudo-differential transmission has been designed in order to reduce coupled noise which is expected to be high, since the distance between the FEE rack and the \NEW\ pressure vessel is about 12 meters. The cable chosen was Twinax TWC-124-2. Placing the coupling capacitor C2 enables the use of the same cable for both supply voltage and signal extraction. In order to achieve a proper TX line matching a $\pi$ type line termination has been used implicit in the amplifier input impedance defined by RA and RB. Both differential impedance of ($120\Omega$) and common mode impedance of ($50\Omega$) have been defined based on TX line characteristics. Common mode termination is of main importance since a pseudo differential connection in the PMT base is being used and common mode signals traveling along the transmission line are expected to be quite high. The FEE board is placed inside a 6U metallic box with grid up and down to allow the air flow. This box is located in the electronic rack near of the acquisition system. HDMI is the physical media chosen for connection between FEE board and ADC as the latter, developed by the RD-51 collaboration long before FEE board was designed, makes use of it due to its differential structure and robustness (see \ref{sec.atca}). A DAC sampling frequency of a decade over FEE signal bandwidth (3 MHz) should be enough to acquire FEE output signals with a neglectable frequency distortion even if low order Nyquist filters are used.
The active components in the FEE have been chosen to minimize noise. It was opted to use differential transmission, since in this way it is possible to cancel the noise coupled to the signal. The differential amplifier chosen is the THS4511 from Texas Instruments, which is characterized by high bandwidth, low noise and low distortion. After the differential amplification, the signal passes through a passive low-pass filter. Finally, a differential driver (LT6600 from Analog devices) is used that provides sufficient current for the transmission to the external acquisition system. The bandwidth of the FEE (3 MHz) makes it work as a shaping filter, stretching the time length of the single photo-electron response that was provided as a specification by the detector calibration team.

The AC coupling capacitors are 18 nF with a C0G dielectric, which offer excellent properties such as high thermal stability, high ripple current capability, low capacitance change with respect to applied DC voltage and no capacitance decay with time. %Modern C0G (NP0) formulations contain neodymium, samarium and other rare earth oxides. C0G (NP0) ceramics offer one of the most stable capacitor dielectrics available. Capacitance change with temperature is 0 $\pm30ppm/$ $^{\circ}$C which is less than $\pm0.3\%$ $^{\circ}$ from $-55$ $^{\circ}$C to $+125$ $^{\circ}$. Capacitance drift or hysteresis for C0G (NP0) ceramics is negligible at less than $\pm0.05\%$ versus up to $\pm2\%$ for films. Typical capacitance change with life is less than $\pm0.1\%$ for C0G (NP0), one-fifth that shown by most other dielectrics.

The equipment for supplying the high voltage to the PMTs is an universal multichannel power supply system, model SY 1527 of CAEN with the board A1733P:0-4 kV output voltage and < 30 mVpp of voltage ripple.  % and fully control: Programmable parameters for each power channel that include two voltage values (V0set, V1set) and two current limit values (I0set, I1set). The switching from one value to the other is performed via two external (NIM or TTL) input levels (VSEL, ISEL). The maximum rate of change of the voltage (Volt/second) may be programmed for each channel. Two distinct values are available, Ramp-Up and Ramp-Down. 
 
\subsection{Noise Measurements}
In order to check the specifications of the design, noise measurements have been carried out in the laboratory and at the LSC NEXT installation. Measures are related to the data acquisition system (DAQ) least significant bit (LSB). The DAQ offers 12 effective bits over an input range of 2V (LSB = 0.49 mV). No dark count noise effects were taken into account since their effect is neglectable under the experimental conditions. The results are shown in the table \ref{tab:noise}. 

\begin{table*}
	\caption{Noise Measurements (in $LSB_{rms}$)}
	\label{tab:noise}
	
	\begin{center}
		\begin{tabular}{ c c || c | c |}
			& & Laboratory & LSC NEXT \\
			\hline
			\multirow{3}{*}{Direct Measurement} & DAQ & 0.64 & 0.66\\
			& FEE + DAQ & 0.75 & 0.75\\
			& FEE + DAQ + PMT & 0.76 & 0.8\\
			\hline
			\multirow{2}{*}{Indirect Measurement} & FE & 0.39 & 0.36\\
			& PMT & 0.12 & 0.28\\
		\end{tabular}
	\end{center}
\end{table*}

The worst case total noise is 0.8 LSB, which is within specifications. The theoretically computed noise of the FEE was 0.35 LSB without power supply contribution. The indirectly measured FEE noise contribution is 0.36 LSB, very close to specifications. The remaining noise can be attributed to the DAQ system. 

\subsection{Front End Electronics model for simulation}

\begin{figure}
	\begin{center}
		\includegraphics[width=0.5\textwidth]{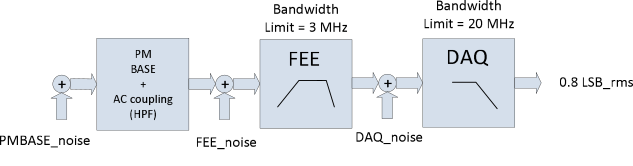}
		\caption{Noise generation scheme.}
		\label{fig:noise_eq}
	\end{center}
\end{figure}

\begin{figure}[!htbp]
	\centering
		\includegraphics[width=0.45\textwidth]{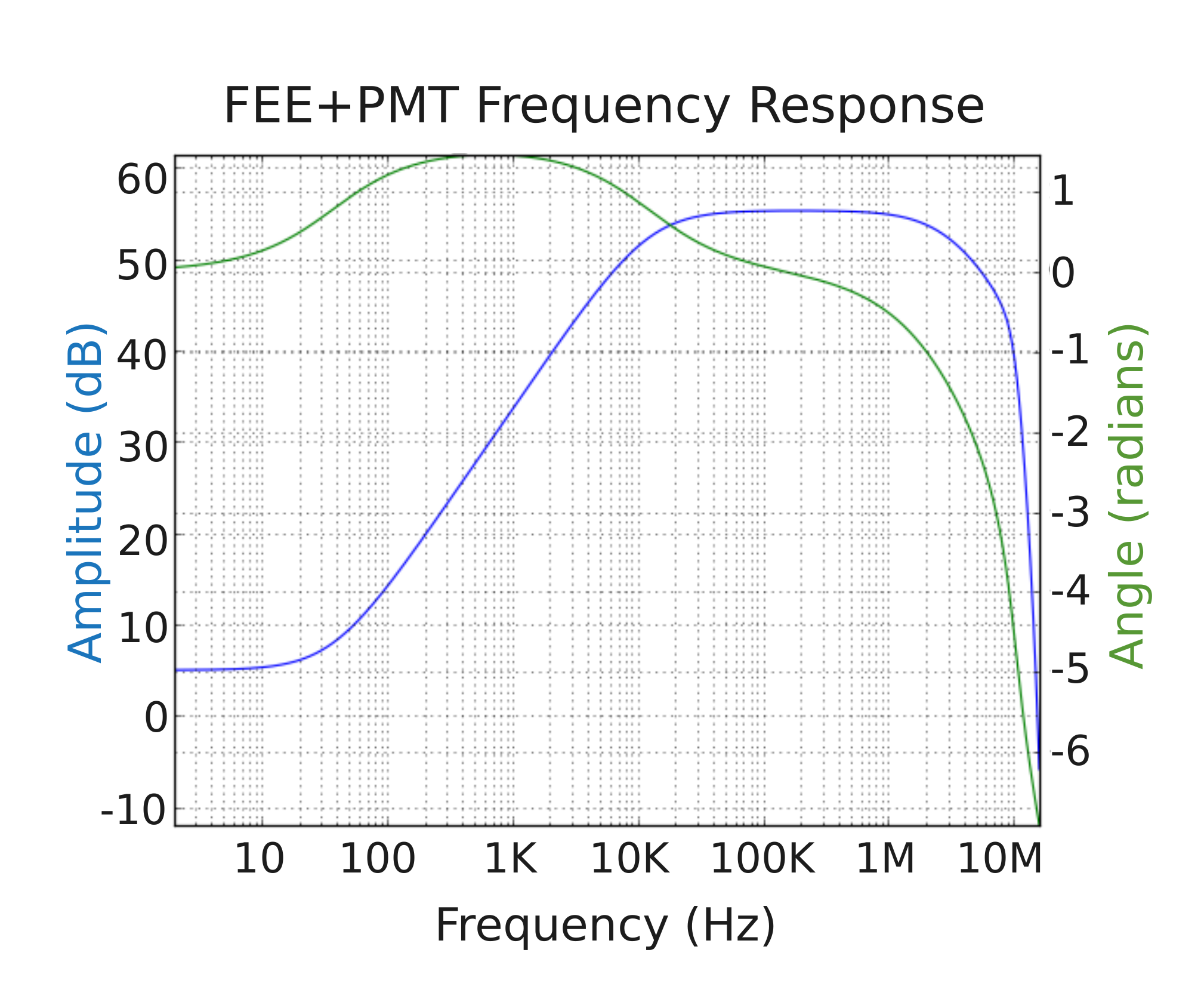}
		\caption{\small FEE Full Frequency Response.}
		\label{fig:Full_Freq}
\end{figure}

The FEE model incorporates the filter described by equation \ref{eq:full_fee} plus a fourth order low pass filter, due to the required shaping (see \fig\ \ref{fig:Full_Freq}). The model response is consistent with detailed SPICE %\cite{SPICE}
simulations of the implemented FEE.

Noise is introduced in the model following the behavior of the real system, that is, taking into account the filtering effect of every part. For instance, the input equivalent noise of the FEE has been increased to show its true effect given the 3 MHz bandwidth. The noise equivalent model of the FEE is shown in \fig\ \ref{fig:noise_eq} and described by the following equations which make use of the noise level measurements obtained in the previous section.

\begin{align}  
GAIN &= FEE_{GAIN}.DAQ_{GAIN} \\ 
vo_{Tn}^{2} &= v_{DAQn}^{2}(out) + v_{F+Pn}^{2}(out) \\
vo_{Tn}^{2} &= \int_{0}^{BW=3MHz}{v_{F+Pn}^{2}.{\lvert}G.H(jw){\rvert}^2}  \\
& + \int_{0}^{BW=20MHz}{v_{DAQn}^{2}.{\lvert}DAQ_{G}.H(jw){\rvert}^2}\\  
vo_{Tn(rms)} &= \sqrt{vo_{Tn}^{2}} = 0.76LSB_{rms}
\label{eq.noise}
\end{align}

Where total gain is the FEE gain multiplied by adquisition system gain. Moreover, the total noise ($vo_{Tn}^{2}$) is the DAQ out noise ($v_{DAQn}^{2}(out)$) plus the noise of the FEE and the PMT base ($v_{F+Pn}^{2}(out)$).

%\begin{eqnarray}
%GAIN &= FEE_{GAIN}.DAQ_{GAIN} \nonumber \\ 
%vo_{TOTAL_{noise}}^{2} &= v_{DAQnoise}^{2}(out) + v_{FEE+PMBnoise}^{2}(out) \nonumber \\
%vo_{TOTAL_{noise}}^{2} &= \int_{0}^{BW=3MHz}{v_{FEE+PMBnoise}^{2}.{\lvert}GAIN.H(jw){\rvert}^2}  \nonumber \\
%& + \int_{0}^{BW=20MHz}{v_{DAQnoise}^{2}.{\lvert}DAQ_{GAIN}.H(jw){\rvert}^2} \nonumber \\
%vo_{TOTAL_{noise(rms)}} &= \sqrt{vo_{TOTAL_{noise}}^{2}} = 0.76LSB_{rms}  
%\label{eq.noise}
%\end{eqnarray}

\section{Baseline Restoration algorithm}
\label{Digi_base_res}

The baseline restoration (BLR) algorithm is based on the implementation of the inverse function of a HPF. %(\Fig\ \ref{fig:FEE_check})
The impulse response of this inverse function has a structure composed of a delta in the origin plus a step function whose amplitude equals the value of $\frac{1}{\tau}$ where $\tau$ is $(R_1+Z_{in}).C$~(equation \ref{eq:imp}). This means that the convolution operation can be carried out using just an accumulator and a multiplier instead of a more complex finite impulse response (FIR) filter.

\begin{eqnarray}
HPF^{-1}(s)&=1+\frac{1}{\tau.s} \nonumber \\
HPF^{-1}(t)&=\delta(t)+\frac{1}{\tau} \int_{0}^{t} dt
\label{eq:imp}
\end{eqnarray}

\begin{figure}[!htbp]
	\centering
	\includegraphics[width=0.35\textwidth]{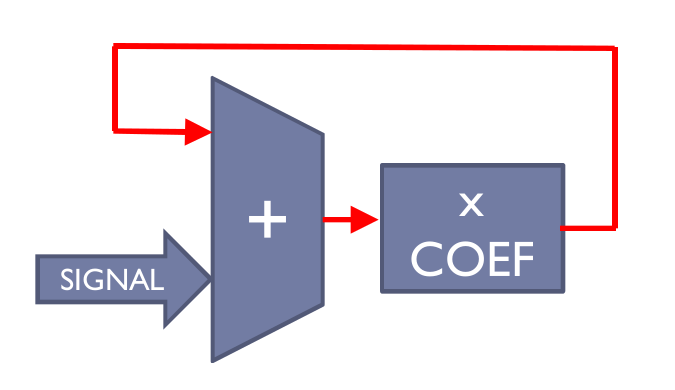}
	\caption{MAC Implementation.}
	\label{fig:FEE_check}
\end{figure}

\begin{figure}[!htbp]
	\centering
		\includegraphics[width=0.45\textwidth]{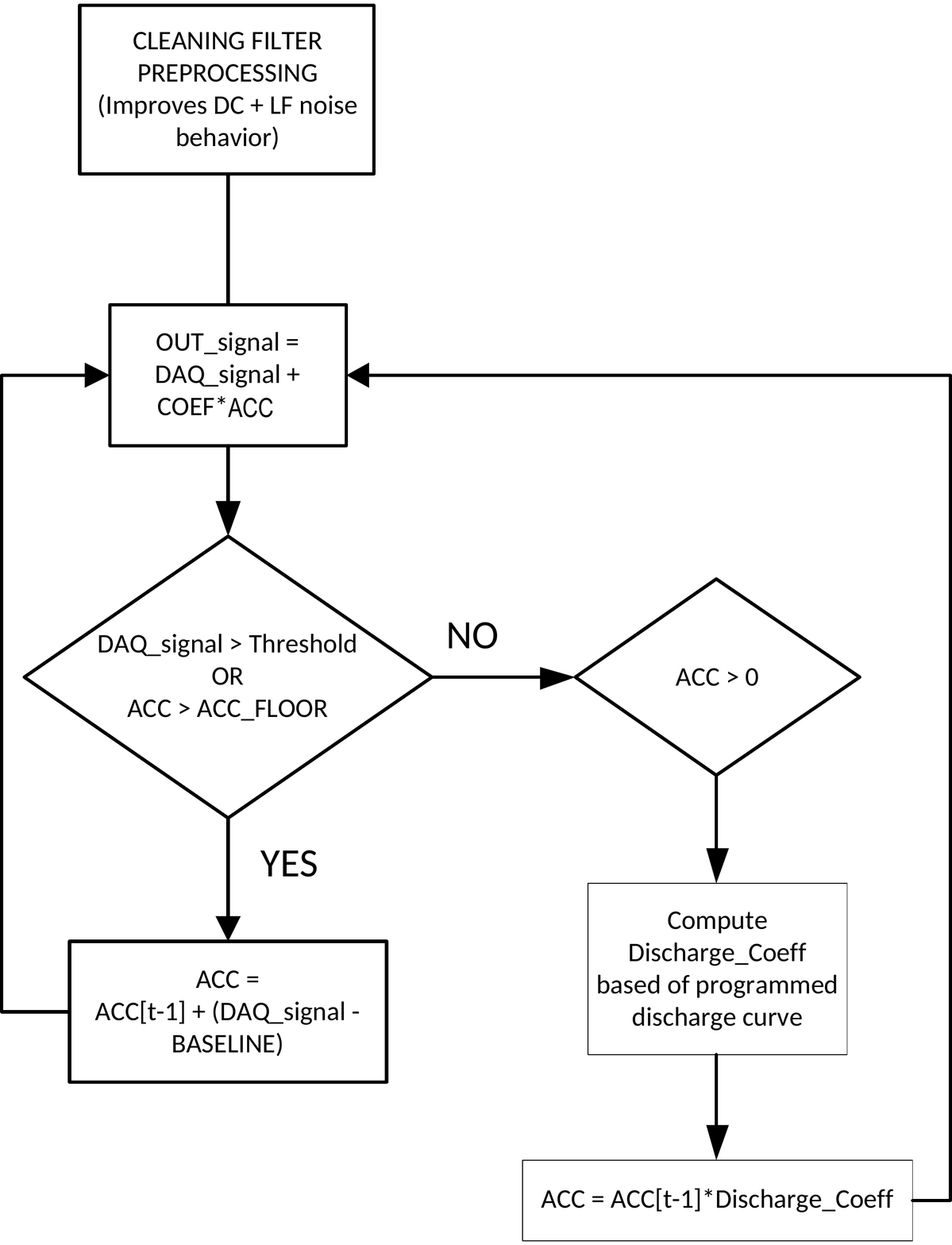}
		\caption{\small Accumulator based BLR.}
		\label{fig:BLR_acc_algor}
\end{figure}

\begin{figure}[!htbp]
	\centering
	\subfloat[Red - Real input signal; Green - FEE output signal]{\includegraphics[width=0.5\textwidth]{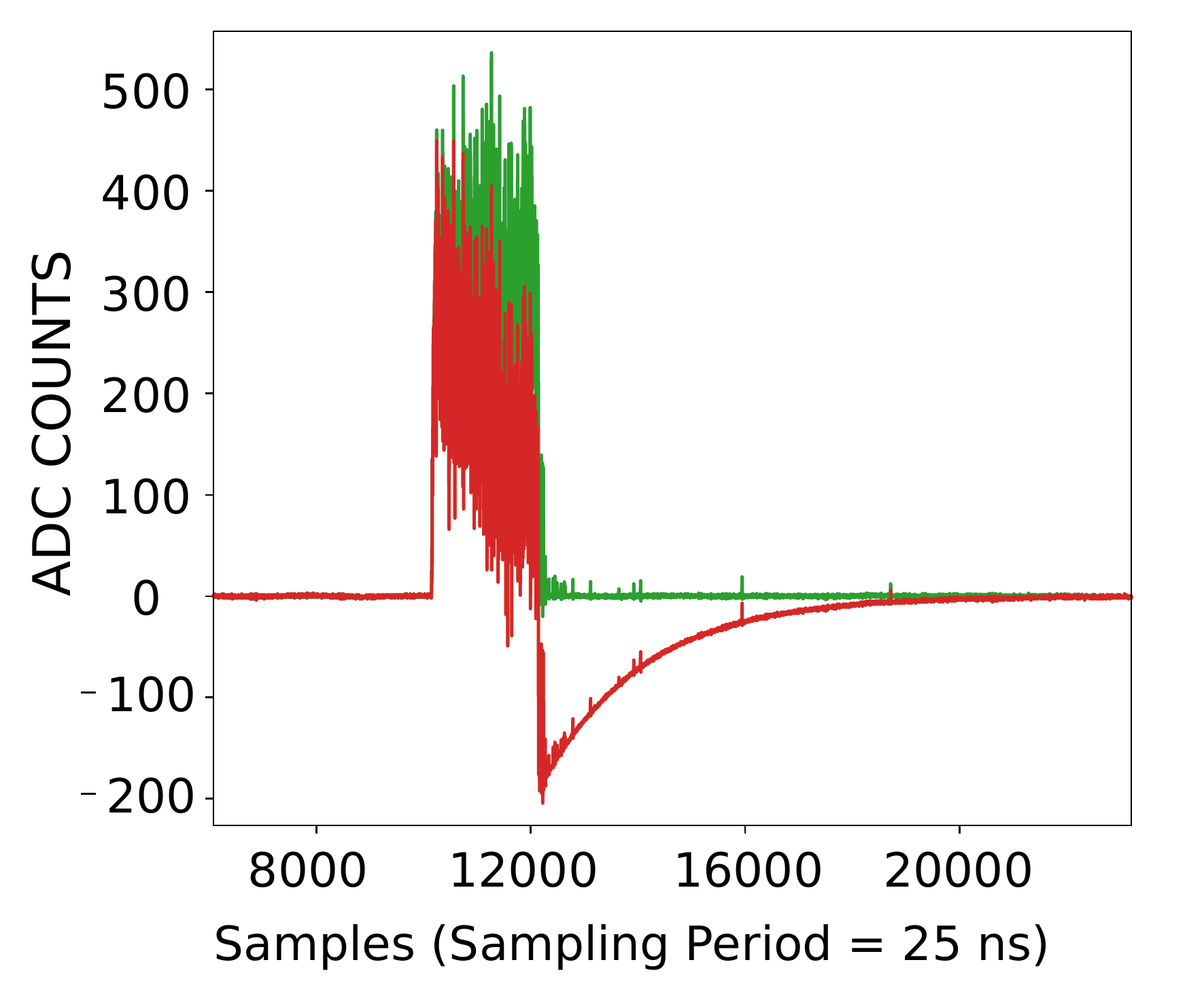}}
	\hfill
	%\subfloat[Blue - Real input signal; Orange - FEE output signal]{\includegraphics[width=0.5\textwidth]{pool/imgEP/signal_preblr_test2.pdf}}

	\subfloat[Digital baseline reconstruction applied]{\includegraphics[width=0.5\textwidth]{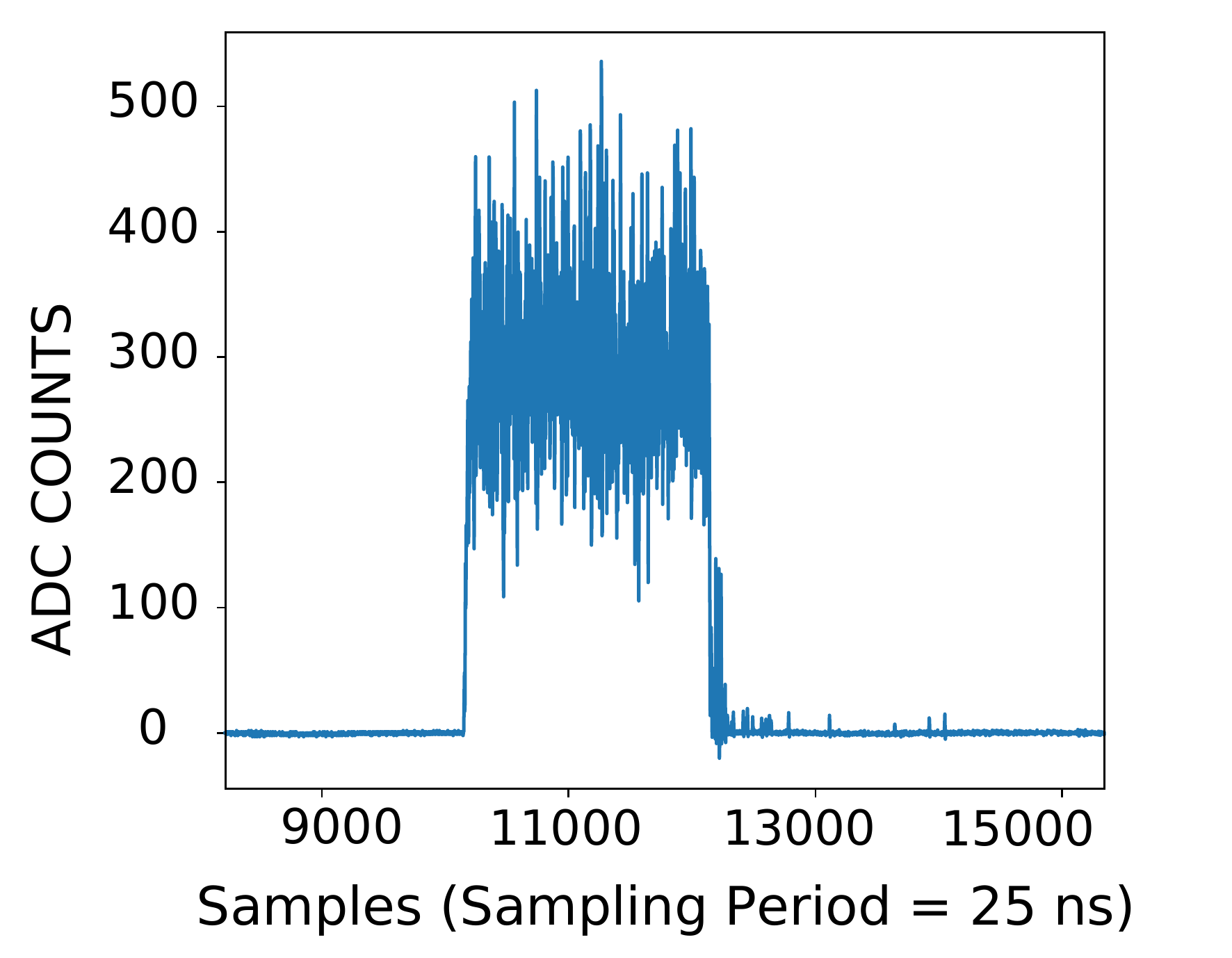}}
	\caption{\small Signal from simulation of 50 $\mu$s pulse. 25865 pe. Noise 0.74 LSB}
	\label{fig:graph_sig}
\end{figure} 

A nice feature of this algorithm is that it can be activated when a pulse is detected and then switched off when the pulse ends. This means that a sizable amount of low frequency noise filtered by the AC coupling capacitor will not be reintroduced in the system, since it is slower than the pulse length.
On the other hand, the low frequency zero introduced by the PMT base interaction adds a low amount of DC to the theoretically AC coupled output signal of the FEE. In order to cancel this effect a high pass filter with a cutoff frequency equal to the frequency of this zero is introduced previous to the BLR algorithm. This completely cancels the DC effect so that the reconstructed signal shows no baseline shift at the end. 

Any algorithm based on a simple accumulator is sensitive to non-zero mean noise. In particular, the residual low frequency components whose components periods are longer than the sampling window (whole signal time window) are prone to leave a small residue inside the accumulator. Also with short pulse signals the different height and length of resulting positive and negative lobes after AC coupling, translates into a non-zero mean error due to quantization effects in the analog to digital conversion. As a consequence the accumulator itself introduces a baseline shift effect due to these non-zero mean noises

The BLR algorithm developed for the \NEW\ energy plane controls the rise of the accumulator using a mechanism to smoothly deplete the residue remaining in the accumulator after a pulse reconstruction. The algorithm flowchart is shown in \Fig\ \ref{fig:BLR_acc_algor}. Notice  that in this algorithm the reconstruction process is started independently from the beginning of the pulse, and can in principle be active continuously. The control relies on the accumulator operations that can be {\em Update} or {\em Discharge}. When the raw signal (coming from the DAQ) rises above a  threshold or the accumulator value is above another threshold (which is the condition for an active pulse) the accumulator is updated as in the original algorithm. However when none of those conditions are met, which means that there is no active signal pulse, the accumulator is forced to a controlled discharge state. This discharge operation is carried out following a smooth curve so that the reconstructed signal shows no jumps or discontinuities. As an example of the action of the BLR algorithm, \Fig\ \ref{fig:graph_sig} shows the output signal of a PMT (showing the characteristic negative swing introduced by the filter) and the corrected signal after the BLR algorithm. 

The estimated residual in the energy correction has been computed applying the deconvolution to Monte Carlo signals generated using the detailed simulation described above and the effect has been quantified to be smaller than 0.3 \% FWHM for long signals (corresponding to large energies), and thus introducing an effect smaller than the resolution due to the Fano factor.

\subsection{Data acquisition system implementation}
\label{sec.atca}

\par In the NEXT experiment Data Acquisition System (DAQ), FPGA-based DAQ modules work in free-running mode, storing data continuously in a circular buffer, while an DAQ Trigger module processes trigger candidates received, generating a trigger accept signal that causes data to be sent to a PC farm. NEXT has adopted the Scalable Readout System (SRS) in the Advanced Telecommunications Computing Architecture (ATCA) for his detector NEW. The SRS-ATCA was defined by the CERN RD51 Collaboration as a multi-channel, scalable readout platform for a wide range of front ends \cite{1748-0221-11-01-C01008}.
 Trigger candidates are generated for each PMT channel in the DAQ Data modules, since each PMT channel is able to sense the primary and/or the secondary scintillation light produced in the chamber. The trigger candidates’ generation is based on the early energy estimation of the events, which requires a stable baseline \cite{1748-0221-7-12-C12001}. 
 As a consequence digital baseline restoration must be implemented online.

\par A similar version of the digital baseline restoration (DBLR) algorithm, introduced in subsection \ref{Digi_base_res} and shown on \fig\ \ref{fig:BLR_acc_algor} has been implemented. This DBLR block is activated whenever the input signal rises above a threshold thus producing an output signal with its baseline completely restored. The threshold  is defined using signal baseline as a reference which requires a precise on-line baseline computation mechanism. A moving average filter has been used in this implementation. In order to avoid baseline shifts due to residues remaining in the accumulator, and further simplify the logic and use of FPGA resources in the DAQ and trigger modules, the control mechanism makes use of a simple linear smoothing function. As in the original algorithm, in case the accumulator is not completely empty when the reconstruction process ends, it must be automatically flushed.

\par The DBLR algorithm has been implemented in a Xilinx Virtex-6 FPGA (XC6VLX240T-1ff1156). There is a DBLR block per channel (12 PMT channels per ATCA-FEC module), and a preceding cleaning filter, both configurable. The implemented algorithm and the cleaning filter have been implemented in a 42-bit fixed point format (Q11.30). The format has been selected as a tradeoff between algorithm stability and physical resources used.

\subsection{Deconvolution of the PMT raw waveforms in \Kr{83m} calibration data}

\NEW\ has been operating at the LSC since October 2016. The detector has been calibrated with data taken with a \Rb{83} source connected to its gas system~\cite{Monrabal:2018xlr}. The exotic rubidium isotope decays to \Kr{83m} via electron capture with a lifetime of \RbLifetime. The krypton then decays to the ground state by emitting two electrons of total energy \textasciitilde 40 keV, with a branching ratio of 95\%. A Kr decay results in an energy deposit of low enough energy to be considered point-like.

%The total released energy sums up to \KrEnergy. A \Kr{83m} decay results, therefore in a point-like energy deposition. 

\begin{figure}[tbh!]
  \begin{center}
    \includegraphics[width=0.45\textwidth]{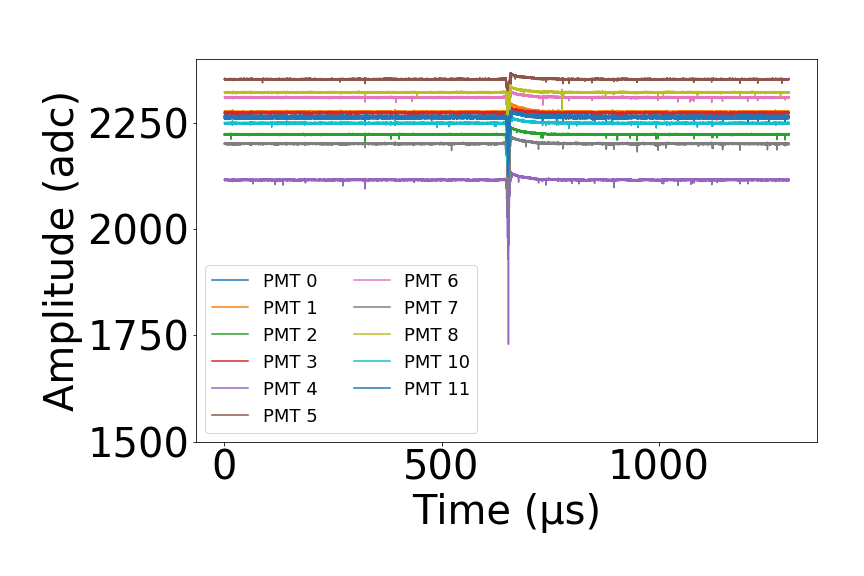} 
    \includegraphics[width=0.45\textwidth]{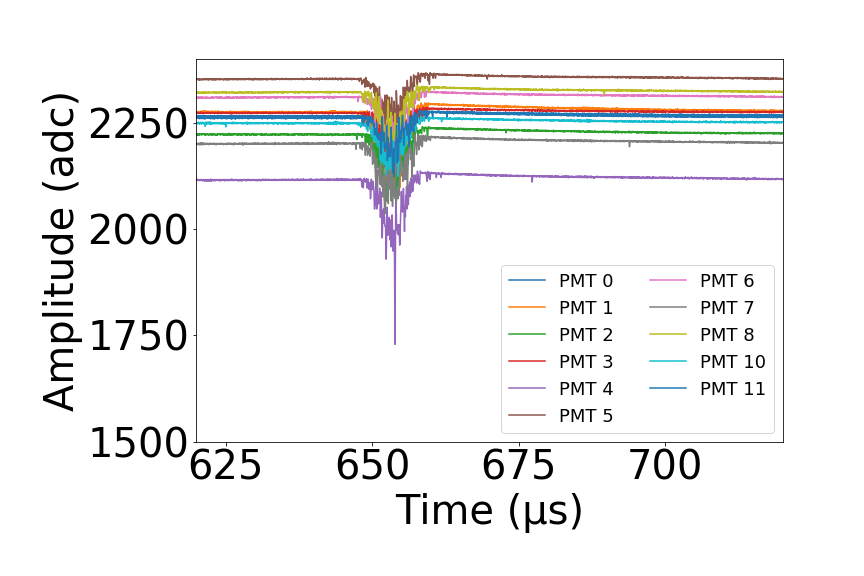} 
    \caption{\label{fig:rwf} \Kr{83m} raw waveforms for the individual PMTs, showing the negative swing introduced by the PMTs front-end electronics. The top image shows the raw waveform in the full data acquisition window, while the bottom image shows a zoom on the EL-amplified signal (\st) on which event read-out was triggered.}
  \end{center}
\end{figure}

\begin{figure}[tbh!]
  \begin{center}
    \includegraphics[width=0.45 \textwidth]{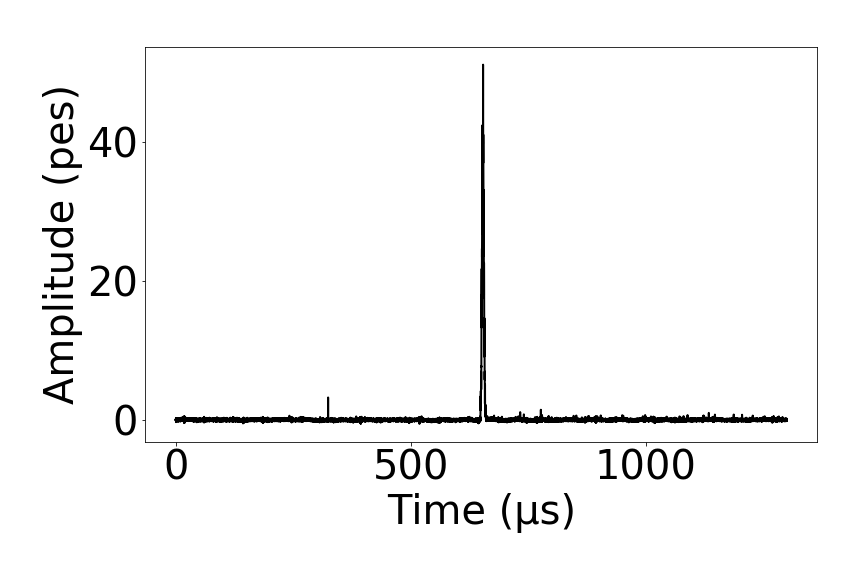} 
    \includegraphics[width=0.45\textwidth]{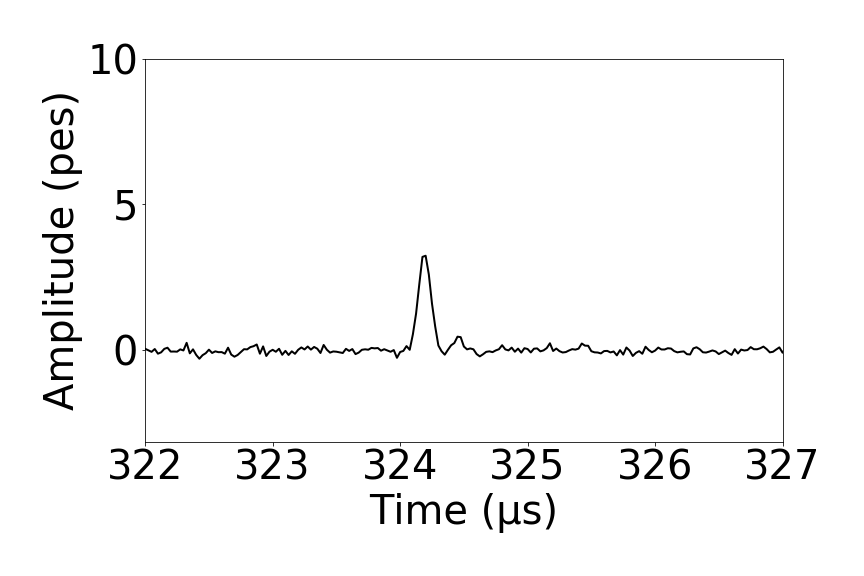}
    \includegraphics[width=0.45\textwidth]{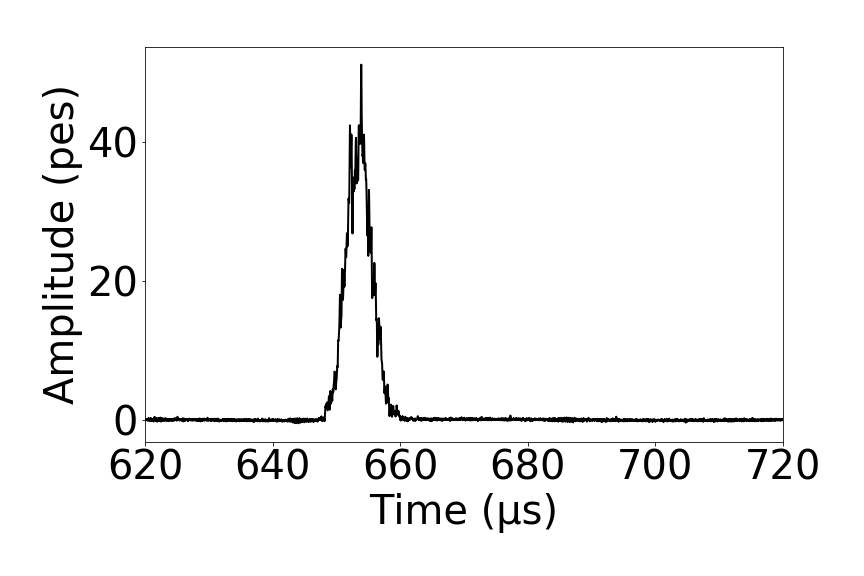} 
    \caption{\label{fig:cwf} \Kr{83m} corrected waveforms for the sum of the PMTs. The top image shows the corrected waveform in the full data acquisition window (\so and \st), while center image shows zooms of the primary scintillation signal (\so) and the bottom image shows EL-amplified signal (\st) waveforms.}
  \end{center}
\end{figure}

The raw data produced by the energy plane are PMT waveforms, sampled each \NewPMTSampling. The HPF associated to the grounded cathode scheme introduces a negative swing in the PMT waveforms. The 
first step in the data processing is to apply the BLR algorithm described above to the arbitrary-baseline, uncalibrated raw-waveforms, to produce positive-only, zero-baseline, calibrated waveforms.  \Fig\ \ref{fig:rwf} shows the raw-waveforms corresponding to the PMTs of the energy plane, while  \Fig\ \ref{fig:cwf} shows the calibrated waveforms corresponding to the PMT sum. The event was triggered by the EL-amplified signal (\st) which appears centered in the data acquisition window. The primary scintillation signal (\so) signal appears at the beginning of the data acquisition window. 

%The deconvolution procedure permits full recovery of the energy of the waveforms. Indeed, the measured energy resolution for \Kr{83m} point-like energy deposits in \NEW\ at \NewSevenBarPressureRunII\ is \ResolutionKrFullFourSevenThreeFourWithSystematics\ (which extrapolates \SQRE\ to \ResolutionKrFullFourSevenThreeFourQbbWithSystematics) in the full chamber and
%\ResolutionKrFidFourSevenThreeFourWithSystematics\ (\ResolutionKrFidFourSevenThreeFourQbbWithSystematics) in a fiducial region
%($\R < \KrFidVolumeRRunII, \Z < \KrFidVolumeZRunII$) chosen to minimize the effect of lower solid angle coverage and large lifetime corrections \cite{Martinez-Lema:2018ibw}.

The deconvolution procedure permits full recovery of the energy of the waveforms. Indeed, the measured energy resolution for \Kr{83m} point-like energy deposits in \NEW\ at \NewSevenBarPressureRunII\ is \ResolutionKrFullFourSevenThreeFourWithSystematics\ FWHM in the full volume. A naive \SQRE\ extrapolation to \Qbb\ yields \ResolutionKrFullFourSevenThreeFourQbbWithSystematics\ averaged over the full volume. A fit in a fiducial region defined as ($\R < \KrFidVolumeRRunII, \Z < \KrFidVolumeZRunII$) has also been evaluated. The radial cut ensures optimal geometrical coverage and the z cut minimizes the residual errors due to lifetime fluctuations, which increase with z. The fit yields \ResolutionKrFidFourSevenThreeFourWithSystematics\, extrapolating to \ResolutionKrFidFourSevenThreeFourQbbWithSystematics\ at \Qbb. This value is reasonably close to the best resolution expected in \NEW\, confirming the excellent capabilities of the technology and the good working conditions of the chamber \cite{Martinez-Lema:2018ibw}.

\section{Conclusion}
\label{sec:conclu} 
A major challenge in the design and implementation of the front-end electronics for the \NEW\ detector's energy plane is to strike the best compromise between the conflicting requirements of radiopurity and performance (linearity, high-gain, etc.) needed for high-resolution measurement of the energy. In this paper, a solution, based on the choice of radio-pure components (Kapton circuits for the PMT bases and low-radioactivity resistors and capacitors), and implementing a grounded cathode PMT connection, has been presented, together with a detailed description of the BLR algorithm which allows a high-precision measurement of the energy. Possible sources of energy resolution degradation in EP have been identified as non-linearity effects in PMTs, bases and FEE, electronic noise (sec \ref{sec:FEE}) and DBLR \ref{Digi_base_res} processing effects. Worst case linearity measurements showed a maximum deviation of 0.38\%, electronic noise has a neglectable effect below 0.8 LSB (0.02\%) and DBLR has been evaluated in MC simulations as a worst case 0.3\% degradation. The combination of all studied effects should degrade energy resolution less than 0.48\%. \NEW\ calibration with \Kr{83m} shows a 0.5 \% energy resolution in the fiducial region which taking into account the intrinsic resolution of 0.3\% would translate into a energy measurement resolution for EP of 0.4\% which is even better than initial estimates.

\section*{Acknowledgments}
We acknowledge support from the following agencies and institutions: the European Research Council (ERC) under the Advanced Grant 339787-NEXT; the Ministerio de Econom\'ia y Competitividad of Spain under grants FIS2014-53371-C04, the Severo Ochoa Program SEV-2014-0398 and the Mar\'ia de Maetzu Program MDM-2016-0692; the GVA of Spain under grants PROMETEO/2016/120 and SEJI/2017/011; the Portuguese FCT and FEDER through the program COMPETE, projects PTDC/FIS-NUC/2525/2014 and UID/FIS/04559/2013; the U.S.\ Department of Energy under contracts number DE-AC02-07CH11359 (Fermi National Accelerator Laboratory), DE-FG02-13ER42020 (Texas A\&M) and de-sc0017721 (University of Texas at Arlington); and the University of Texas at Arlington. We acknowledge partial support from the European Union Horizon 2020 research and innovation programme under the Marie Sklodowska-Curie grant agreements No. 690575 and 674896. We also warmly acknowledge the Laboratorio Nazionale di Gran Sasso (LNGS) and the Dark Side collaboration for their help with TPB coating of various parts of the \NEW\ TPC. Finally, we are grateful to the Laboratorio Subterr\'aneo de Canfranc for hosting and supporting the NEXT experiment.

\section*{References}
\bibliographystyle{NextRefsStyle}
\bibliography{pool/NextRef}
\end{document}